\documentclass[prb,aps,twocolumn]{revtex4}

\usepackage{graphicx}

\usepackage{dcolumn}

\usepackage{bm}

\begin{document}

\title{Electrochemical integration of graphene with light absorbing copper-based thin films}

\author{Medini\ Padmanabhan,$^1$ Kallol\ Roy,$^1$ Gopalakrishnan Ramalingam,$^2$ Srinivasan Raghavan,$^2$ and Arindam\ Ghosh$^1$}

\address{$^1$Department of Physics, Indian Institute of Science, Bangalore 560012, India}
\address{$^2$Materials Research Center, Indian Institute of Science, Bangalore 560012, India}

\date{\today}

\begin{abstract}

We present an electrochemical route for the integration of graphene with light sensitive copper-based alloys used in optoelectronic applications. Graphene grown using chemical vapor deposition (CVD) transferred to glass is found to be a robust substrate on which photoconductive Cu$_{x}$S films of 1-2 $\mu $m thickness can be deposited. The effect of growth parameters on the morphology and photoconductivity of Cu$_{x}$S films is presented. Current-voltage characterization and photoconductivity decay experiments are performed with graphene as one contact and silver epoxy as the other.

\end{abstract}


\maketitle

\section{Introduction}

Carbon based nanostructures such as nano-tubes (CNTs) and graphene are starting to carve out a niche for themselves in the field of energy related materials. Their applications range from hydrogen storage architectures to design of ultracapacitors \cite{dimitrakakisNL08,yuJPCL10}. In the field of optoelectronics both CNTs and graphene are being regularly incorporated in a wide variety of solar photovoltaic designs. For CNTs, the strategy has been twofold: (1) direct employment as photosensitive material \cite{kongkanandNL07,dissanayakeNL11}, and (2) as transparent conducting films for front electrodes \cite{wuSci04,pasquierAPL05,barnesAPL07}. Both strategies have been investigated for graphene as well. Functionalized (organic solution processed) graphene has been demonstrated to be efficient electron acceptors in organic bulk heterojunction photovoltaic devices \cite{liAdMa11}. Solar cell designs which have successfully integrated graphene as the transparent electrode including silicon Schottky junctions \cite{liAdMa10,chenNL11}, dye-sensitized cells \cite{wangNL08,hongEChComm} as well as thin film organic \cite{wuAPL08} and inorganic \cite{biAdMa11} cells, seem to have generated particular interest.

Current market for transparent electrodes is dominated by indium tin oxide (ITO) thanks to a rather rare combination of high transparency and conductivity. However, material scarcity, brittleness, and operational deterioration due to ion diffusion have been some of the outstanding problems \cite{schlatmannAPL96,chenEFM02}. Graphene is an especially attractive alternative for such applications due to its compatibility with planar technology, superior mechanical qualities, transparency and reasonably high conductivity. The development of chemical vapor deposition (CVD) of graphene on transition metals, such as copper and nickel \cite{liSci09}, has made large-area scalability plausible \cite{baeNnano10}. There are however several challenges to this, in particular, (1) creating high specific conductance that can be comparable to ITO, and (2) implementing a robust and non-invasive technique to transfer and integrate graphene to the light absorbing components of the solar cell. While significant progress has been made to produce high-conductivity CVD graphene \cite{wangAdMa11,liJACS11}, its integration to photosensitive components is still in a rather primitive stage. The conventional way for incorporating graphene to solar cell architectures is to lay a sheet of graphene on the active layers \cite{liAdMa10,chenNL11}. This can trap significant quantity of impurities, water/gas molecules, acrylic residues etc. Thus a clean and non-invasive method of integrating graphene to photovoltaic designs may not only enhance the yield and efficiency, but also create a platform to realize a larger class of hybrid energy-harvesting structures.

\begin{figure}
\includegraphics[scale=1]{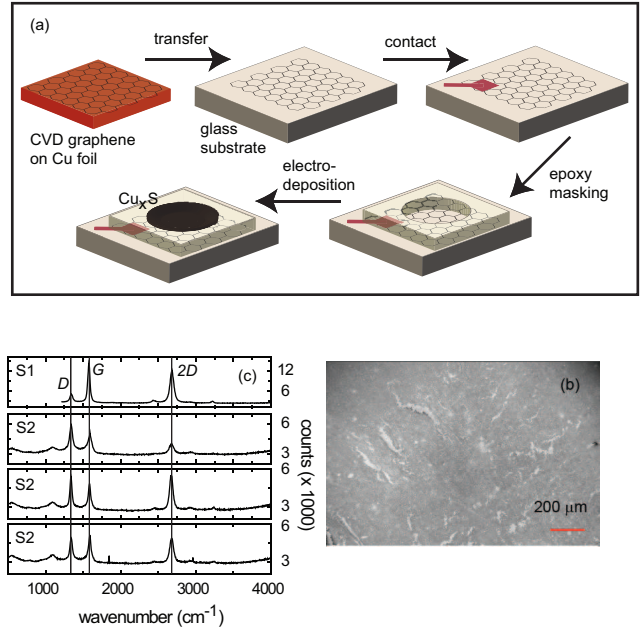}
\caption{(a) Process flow diagram for the electrochemical integration of CVD graphene with photosensitive materials (b) SEM image of graphene transferred to glass  (c) Raman spectra of graphene on glass. Spectra from two samples are given (S1 and S2). The three traces for sample S2 correspond to three different positions of the incident LASER beam.}
\end{figure}

Here we demonstrate a new room-temperature integration scheme (Fig. 1(a)) of single-layer CVD graphene to copper-based inorganic photosensitive alloys through electrochemical means. We show that graphene can be used as a cathode on which copper-based
alloys, in this case copper sulfide (Cu$_x$S), can be directly grown from electrochemical bath of appropriate salts. Preliminary electro-optical characterization of the devices is reported and compared with identically grown ITO-Cu$_x$S devices. Our results suggest that graphene can be an excellent host in electrochemical coupling to a wide variety of alloys and materials.

\section{Results and discussion}

The starting point for the preparation of our samples is graphene synthesized by low-pressure CVD (base pressure of 1 Torr) \cite{palACSnano11}. 25 $\mu$m thick copper foils are annealed at 1000 $^{\circ}$C for 5 minutes under a H$_2$ flow of 50 sccm (standard cubic centimeters per minute) to reclaim the pure metal surface. CH$_4$ and H$_2$ are then introduced at a rate of 35 sccm and 2 sccm for a growth time of 30 s. The reactor is cooled down to room temperature at a cooling rate of 8$^{\circ}$C/minute under a 1000 sccm flow of H$_2$. Scanning electron microscope (SEM) images of the as-grown graphene shows complete coverage of the copper substrate. The SEM images also suggest that the growth is predominantly single-layer \cite{vidyaJAP11}.

Small pieces (5 mm $\times$ 5 mm) are cut out of these copper foils and then transferred to a petri-dish containing acidic FeCl$_3$ solution which etches away the copper. The graphene which floats on the surface of FeCl$_3$ is then scooped and transferred into another petri-dish containing de-ionised water. It is allowed to float for about 2$-$3 minutes before being scooped on to a piece of clean glass. Note that no PMMA (poly(methyl methacrylate)) is used in our transfer procedure. This is because electrodeposited Cu$_{x}$S is observed to have poor adhesion on graphene films transferred with the help of PMMA. We believe that PMMA residues on the surface of graphene are responsible for this. Ar-H$_{2}$ annealing may be helpful in improving the interface properties of graphene transferred using PMMA.

In Fig.\ 1(b), we show SEM images of graphene transferred to glass substrates using the above procedure. Note that large area transfer is possible with a few tears. In Fig.\ 1(c), we show representative Raman spectra taken on our samples with 514 nm radiation. In most of the cases, the $D$-peak is stronger or comparable to the $G$ peak indicating high level of disorder. Note that our transfer procedure can cause ripples and tears resulting in lattice distortions which might be the reason for the prominent $D$-peak. Our $2D$ peaks are well-fitted by a single lorentzian, thereby indicating that our substrates are predominantly single layer \cite{ferrariSSC07}. Measurement of the sheet resistance of our transferred graphene sheets using van der Pauw technique yields values between 5$-$15 $k\Omega/sq$. The presence of micro-tears in our sample makes van der Pauw measurements difficult \cite{pauwPRR58}. More careful Hall bar measurements done using similar graphene samples transferred to clean Si/SiO$_2$ substrates yield resistivity values of the range 5$-$50 k$\Omega$. Although these values are rather high, they can be brought down by various methods including chemical doping and gating \cite{wadhwaNL10,jiaNL11}.

\begin{figure}
\includegraphics[scale=1]{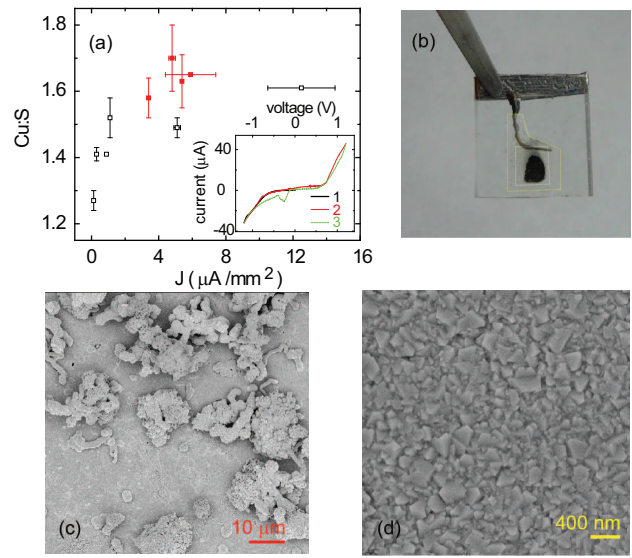}
\caption{ (a) Copper to sulphur ratio (Cu:S) of Cu$_{x}$S films as a function of growth current density. The red (closed) symbols indicate samples which are photoconductive. inset: $I-V$ characteristics of the electrode in the electrochemical bath. Data from three consecutive sweeps are shown as indicated by the numbers. The first sweep is from 0 V to -1.2 V while the second (third) is an up- (down-) sweep between -1.2 V and +1.2 V. (b) Photograph of the substrate along with the grown Cu$_{x}$S film held by tweezers. The outer boundaries of the epoxy mask and the underlying graphene are indicated by the dotted lines. The glass piece is of a lateral dimension of $\sim$1 cm (c) SEM image of a film grown with a current density of 3 $\mu$A/mm$^{2}$. (d) Microstructure of the under-layer of the film shown in (c).  }
\end{figure}

We use graphene transferred onto glass as the substrate for electrochemical deposition. We choose copper sulfide as our light absorbing material because it has an indirect bandgap of about 1.2 eV (for Cu$_{2}$S) and the component elements are earth abundant and non-toxic \cite{wuNL08}. Electrochemical routes for the synthesis of copper sulfide on metal substrates are also well-researched \cite{yukawaTSF96,maneMCP00,anuarSEMSC02}. Solar cells with efficiencies of around 10$\%$ have been reported for the Cu$_2$S-CdS system \cite{hallTSF79,bragagnoloIEEE80}. Recently there has been a revival of interest in the electrodeposition of related materials such as CuInSe$_2$ and Cu(In,Ga)Se$_{2}$ \cite{valdesEChActa11,ribeaucourtEChActa11}.

We now describe our electrochemical growth procedure. The graphene transferred to glass is contacted using silver epoxy paste. For electrochemical deposition, it is desirable that only a pre-defined area is exposed to the growth solution. This is achieved by masking graphene by non-conducting epoxy. This epoxy is also found to be helpful in clamping the graphene firmly to glass. The electrochemical bath consists of 10 mM CuSO$_{4}$.5H$_{2}$O (10 mmoles of the solute for every 1L of the solvent), 400 mM Na$_{2}$S$_{2}$O$_{3}$.5H$_{2}$O and 24 mM EDTA (dihydrate) disodium salt. The freshly prepared solution is aged for about 12 hours maintaining a pH of around 3.0 $\pm$ 0.2. The pH is lowered to $\sim$2.8 right before the growth starts and no further pH maintenance is done during growth. We observe that the pH slowly rises to about 3.5 during a course of 7 hours which is our typical growth time. Graphite is used as the counter electrode in our two-electrode set-up. Typical current densities and voltages for deposition are 3$-$5 $\mu$A/mm$^{2}$ and 0.9$-$1.1 V respectively.

In the inset of Fig. 2(a) we show the $I-V$ characteristics of the electrode at a voltage sweep rate of 5.5 mV/s. Data for three sweeps are given. From the third sweep onwards, a peak starts developing around -0.25 V during downsweeps which we believe is indicative of a reaction involving copper. We observe that even after repeated sweeps ($\sim$50) the graphene substrate is found to be unaffected (under optical microscope) indicating that the deposition process is reversible. For electrodeposition, a constant negative bias is applied to the graphene electrode (cathode). In the electrolyte, Cu$^{2+}$ ions are present due to the dissociation of CuSO$_{4}$. Both thiosulfate and EDTA act as complexing agents for copper. The thiosulphate ion is known to reduce Cu$^{2+}$ to Cu$^{+}$. The sulphur source is the thiosulphate radical which releases colloidal sulphur in acidic medium \cite{yukawaTSF96,anuarSEMSC02}. A photograph of the grown film is shown in Fig. 2(b). We observe that the adhesion of the copper sulfide film to graphene is strong in spite of the underlying tears and disorder in graphene.

\begin{figure}
\includegraphics[scale=1]{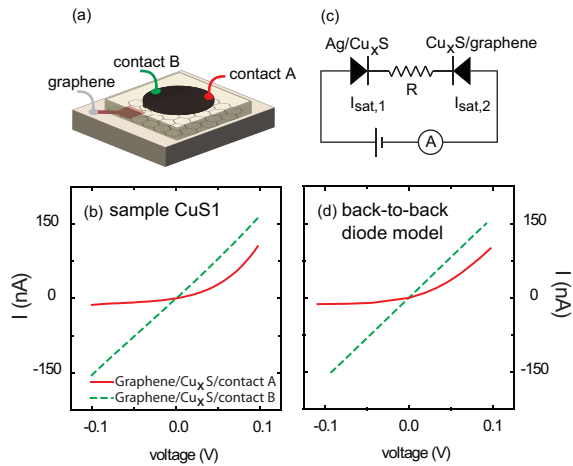}
\caption{(a) Schematic representation of the contact configuration. The two Ag-epoxy contacts on Cu$_{x}$S (A and B) are different in terms of their areas and possibly, Schottky barrier heights. (b) Experimental two-point $I-V$ characteristics observed in our samples with graphene and silver epoxy as the two terminals. (c) Back-to-back diode model used to understand the experimental data. (d) Analytical curves obtained from the model. The parameters for the almost linear curve are $I_{sat,1} = I_{sat,2} = 5.0 \times 10^{-7}$ A and R = 500 k$\Omega$. The parameters for the non-linear curve are $I_{sat,1} = 1.3 \times 10^{-8}$ A, $I_{sat,2} = 4.0 \times 10^{-7}$ A and R = 300 k$\Omega$}
\end{figure}

It is known that copper forms a series of sulfides with varying ratios of Cu:S. For photovoltaic applications, copper rich phases, mainly Cu$_{2}$S, are the preferred ones \cite{bragagnoloIEEE80,hallTSF79,caswellJPD77}. In this study, we typically grow films with Cu:S ratios of 1.3 - 1.7. The composition of our films is determined by energy dispersive spectroscopy (EDS). Figure 2(a) shows measured Cu:S ratios as a function of growth current density. At lower current densities (0.1 - 1 $\mu$A/mm$^{2}$), the coverage on the substrate is sparse and the Cu:S ratio is typically low ($\lesssim$1.4). Photoconductivity is not generally observed in these films. This is expected since many of the copper-poor phases are known to be metallic. As the growth current density is raised, the coverage on the substrate improves and complete coverage is achieved by current densities more than $\sim$3 $\mu$A/mm$^{2}$. In Fig. 2(c) we show a film grown near the optimum current density. An underlayer of Cu$_{x}$S is visible, a magnified image of which is shown in Fig. 2(d). A copper-rich overlayer growth is also typically seen which becomes more pronounced at higher current densities ($\sim$10 $\mu$A/mm$^{2}$). At high current densities, a co-deposition of elemental copper is also observed in the overlayer. However, we believe that it is the underlayer that contributes to the measured photoconductivity. As can be inferred from Fig. 2(a), a copper-rich underlayer is conducive to the observation of photoconductivity. Further studies have indicated that a copper-rich underlayer can be obtained by tuning various other parameters including the relative concentration of the bath elements, pH, the presence of complexing agents and deposition voltage.

Ideally, during electrodeposition, we expect the copper sulfide film to grow vertically on top of graphene. However, in our experiments we observe that, the film also creeps up laterally on top of the underlying non-conducting epoxy mask. In many cases, the microstructure and composition of the lateral growth is observed to be comparable to the underlayer of the main film. For electrical measurements we exploit lateral growth to make contacts to the sample. Attempts to directly contact the thin film from above resulted in short-through to the underlying graphene in some cases.

In Fig. 3(a) we show a schematic of the contact configuration in our devices. We take silver epoxy contacts from the underlying graphene and the laterally grown copper sulfide. Examples of two-point $I-V$ characteristics are shown in Fig. 3(b). The two traces correspond to two different silver epoxy contacts (A and B) on the same sample. These two contacts differ from each other in terms of their areas and possibly, Schottky barrier heights. The graphene contact used is same for both the traces. Note that the shape is linear in one case, whereas it is rectifying in the other. In general, the range of currents allowed through the different contact pairs also vary widely.

We try to understand the shape of these curves and the  magnitude of the currents by using a back to back diode model as shown in Fig. 3(c). Ideal diode equation is used, $I = I_{sat}[\exp(eV/nkT)-1]$, where $I$ is the current through the diode, $V$ is the voltage drop across the diode, $I_{sat}$ is the reverse saturation current, $kT$ is thermal voltage and $n$ is the non-ideality factor assumed to be 1.1 in our case. In Fig. 3(d), we show examples of computed curves corresponding to this back-to-back diode model. Note that results comparable to our experiential curves can be reproduced by choosing appropriate values of $I_{sat}$ and $R$. We compare curves taken under various contact configurations and conclude that our results are best explained by assuming that the Cu$_{x}$S layer is n-doped.

We repeat $I-V$ measurements using two Ag/Cu$_x$S junctions as contacts and conclude that they are non-linear in character. The nature of the graphene/Cu$_x$S junctions, however, is unclear. Within our model, it is difficult to distinguish between an ohmic contact and a Schottky contact with a high reverse saturation current. Note that the area of graphene junction is much larger than the silver junction which will lead to high $I_{sat}$. We have devices where as much as 1 $\mu$A current flows at 0.1 V when the graphene junction is reverse biased. This can indicate a Schottky barrier with $I_{sat}\gtrsim$ 1 $\mu$A or an ohmic contact with resistance of $\sim$100 k$\Omega$. However, in most cases, the resistance of the Cu$_{x}$S film is present in series with the junction resistances. These series resistances are typically large, probably due to the fact that the laterally grown material has uneven substrate coverage and is prone to micro-cracks.

In Fig. 4(a) we show $I-V$ characteristics of our device under dark and light (AM1.5, 1000 W/m$^{2}$) conditions. An anti-clockwise rotation of the curve is observed in response to light which indicates a decrease in resistance. In this trace one contact is graphene and the other one is Ag-epoxy on laterally grown Cu$_{x}$S. Two point measurements with a different contact configuration where both contacts are taken from laterally grown Cu$_{x}$S also show similar light response. Since the contact area of graphene is much larger than Ag-epoxy, we conclude that the magnitude of the rotation of the $I-V$ curves in response to light is mostly independent of the contact area. Hence, we believe that major contribution to the observed photoconductivity comes from the bulk of the sample and not just the region surrounding the contacts.

\begin{figure}
\includegraphics[scale=1]{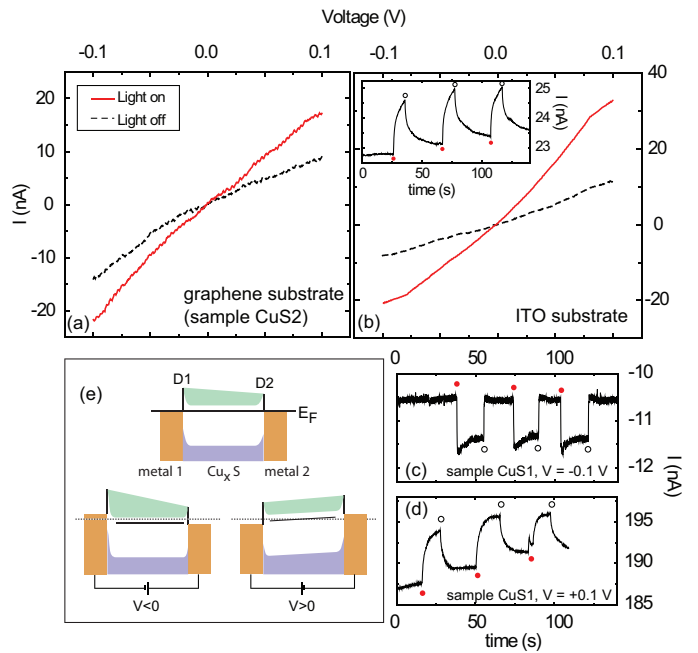}
\caption{(a) $I-V$ characteristics of Cu$_{x}$S film grown on graphene under dark and light (AM1.5) conditions (b) Similar characteristics for the film grown on ITO. inset: current vs time taken at a constant voltage of +0.1 V. An LED is turned on and off periodically as shown by the red (closed) circles (light on) and black (open) circles (light off). (c) and (d) Photoresponse of the Cu$_{x}$S film grown on graphene at V = -0.1 V and +0.1 V, respectively. (e) Schematic showing the voltage drop across various segments of the device as a function of the polarity of the external bias. }
\end{figure}

In Fig. 4(b), similar data is shown for a film grown on ITO coated glass. This film was grown with a current density of 3 $\mu$A/mm$^{2}$. Two point measurements are done with ITO as one contact and Ag-epoxy contacted to laterally grown Cu$_{x}$S as the other. Note that the responses of both films to light are comparable in spite of the fact that the sheet resistance of ITO is only $\sim$10 $\Omega/sq$. However, the difference between ITO and graphene may be masked by the fact that the overall resistance has a significant contribution from the thin film itself.

In Figs. 4(c) and 4(d) we monitor the current through the films grown on graphene as a function of time. This data is shown for sample CuS1, whose $I-V$ characteristics are shown in Fig. 3(b) (graphene/Cu$_{x}$S/contact A). A constant bias of $\pm$0.1 V is maintained. A white light emitting diode (LED) is turned on and off periodically and the response is recorded. In the inset of Fig. 4(b), we show photoconductive build-up and decay measured for the Cu$_{x}$S film grown on ITO substrates. We observe that the response to light evolves over timescales of tens of seconds. It is known that traps present in the bulk and interfaces cause slow decay of photoconductivity by trapping the minority carries for long timescales thereby delaying their recombination with the majority carriers. In some cases, the shape of our photoconductive decay is well-fitted by a stretched exponential. The presence of DX centers and local-potential fluctuations due to compositional inhomogeneities are possible factors which can influence the decay \cite{linPRB90,linPRB90b}.

Note that in Figs. 4(c) and 4(d) the only difference is the sign of the applied bias. The shape of the decay curves are however markedly different. The general observation is that the photoresponse curves are dominated by a fast time scale when certain contacts with low values of $I_{sat}$ are reverse biased. We try to understand the bias dependence of the light response timescales by considering a model in which copper sulfide is assumed to be an n-type semiconductor with two unequal Schottky barriers at two contacts (Fig. 4(e)). The junction at the left (D1) is assumed to have a lower value of $I_{sat}$ compared to the junction at the right (D2). When a negative bias is applied to D1, most of the external voltage is dropped across the depletion region of D1. The timescales of the photoresponse is thus dictated by the processes in the depletion region. However, when the sign of the bias is reversed, the voltage drops across D1 and D2 are comparable due to the forward bias and large value of $I_{sat,D2}$ respectively. The timescales in this case can have significant contribution from the processes in the bulk. However, if the two junctions have comparable values of $I_{sat}$, the device becomes symmetric. In such cases, our experiments show that the light response curves at positive and negative biases have qualitatively similar features.

Further studies including temperature and bias dependent transport measurements are needed before we can fully understand the processes in our system. Local current density variations during electrochemical growth owing to the high sheet resistance of graphene as well as presence of impurities can cause compositional inhomogeneities in the film which can also influence transport data in the present device configuration. For optoelectronic applications, it is also desirable to anneal the films before characterizing them since electrodeposition is known to introduce considerable disorder.

\section{Conclusion}
In conclusion, we present a novel method of integrating graphene with photovoltaic device architectures. We employ an electrochemical route to grow a thin film of Cu$_{x}$S on CVD graphene and investigate its electrical and optical properties. Cu(In,Ga)(S,Se)$_{2}$ and Cu$_{2}$ZnSn(S,Se)$_{4}$ which are known to be near-ideal materials for solar cell applications, also fall into the same family of chalcogenides \cite{bhattacharyaSEMSC03,lincotSE04,fernandezTSF05,katagiriTSF09,todorovAdMa10}. Recently, electrochemical growth of CuInSe$_2$ on CNT-based nanocomposite membranes has been reported \cite{kouEChActa2011}. The device architecture outlined in our work has many advantages including simplicity, low-cost and scalability thereby opening up new avenues for integration of graphene with various optoelectronic devices including solar cells.

\section{Acknowledgement}
We thank Korean Institute of Science and Technology for funding and Prof. V. Venkataraman for illuminating discussions. We acknowledge the Department of Science and Technology (DST) for a funded project. S.R. acknowledges support under Grant No. SR/S2/CMP-02/2007.

\bibliographystyle{achemso}
\bibliography{refs}

\end{document}